\documentclass[aps,prb,reprint,amsmath,amssymb,groupedaddress]{revtex4-2}
\usepackage{graphicx}% Include figure files
\usepackage{dcolumn}% Align table columns on decimal point
\usepackage{bm}% bold math

\begin{document}

%\preprint{}

\title{Commensurability oscillations in the Hall resistance of unidirectional lateral superlattices}


\author{Akira Endo}
\email[]{akrendo@issp.u-tokyo.ac.jp}
%\thanks{}
\affiliation{The Institute for Solid State Physics, The University of Tokyo, 5-1-5 Kashiwanoha, Kashiwa, Chiba 277-8581, Japan}

\author{Shingo Katsumoto}
\affiliation{The Institute for Solid State Physics, The University of Tokyo, 5-1-5 Kashiwanoha, Kashiwa, Chiba 277-8581, Japan}

\author{Yasuhiro Iye}
\affiliation{The Institute for Solid State Physics, The University of Tokyo, 5-1-5 Kashiwanoha, Kashiwa, Chiba 277-8581, Japan}


\date{\today}

\begin{abstract}
We have observed commensurability oscillations (CO) in the Hall resistance $R_{yx}$ of  a unidirectional lateral superlattice (ULSL). The CO, having small amplitudes ($\sim$ 1 $\Omega$) and being superposed on a roughly three orders of magnitude larger background, are obtained by directly detecting the difference in $R_{yx}$ between the ULSL area and the adjacent unmodulated two-dimensional electron gas area and then extracting the odd part with respect to the magnetic field. The CO thus obtained are compared with a theoretical calculation and turn out to have the amplitude much smaller than the theoretical prediction. The implication of the smaller-than-predicted CO in $R_{yx}$ on the thermoelectric power of ULSL is briefly discussed.
\end{abstract}

%    insert suggested PACS numbers in braces on next line
%    \pacs{73.43.Qt, 73.23.-b, 73.21.Cd}

%\keywords{}

\maketitle

\section{Introduction\label{SecIntr}}

\begin{figure}
\includegraphics[width=8.6cm,clip]{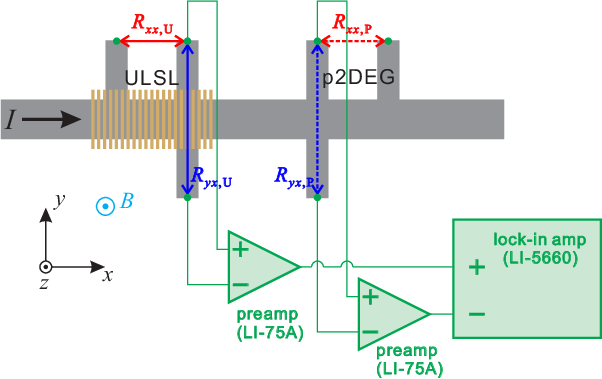}%
\caption{Schematic drawing of the Hall bar device containing, in series, areas with (ULSL) and without (p2DEG) the periodic potential modulation $V(x)$.  Wiring for directly measuring the excess Hall resistance, $\Delta R_{yx} = R_{yx,\mathrm{U}} -R_{yx,\mathrm{P}} $, introduced by $V(x)$ is also shown. \label{sampleschem}}
\end{figure}

Commensurability oscillations (CO), also known as Weiss oscillations, have been arguably one of the best-known magnetoresistance phenomena in mesoscopic systems since their discovery in 1989 \cite{Weiss89,Winkler89}. They were uncovered in a unidirectional lateral superlattice (ULSL), a two-dimensional electron gas (2DEG) subjected to a weak one-dimensional (1D) periodic modulation $V(x)$ of the electrostatic potential. %, and arise from the commensurability between the cyclotron radius $R_\mathrm{c}$ and the period $a$ of the modulation. 
The most prominent oscillations were observed in the magnetoresistance $R_{xx}$ along the modulation, with the minima taking place at the flat-band conditions,
\begin{equation}
\frac{2 R_\mathrm{c}}{a} = n - \frac{1}{4},\hspace{12mm} (n = 1, 2, 3, ...), \label{flatband}
\end{equation}
where $a$ is the period of $V(x)$ and $R_\mathrm{c} = \hbar k_\mathrm{F}/(e|B|)$ is the cyclotron radius, with $k_\mathrm{F} = \sqrt{2 \pi n_e}$ the Fermi wavenumber, $n_e$ the electron density, and $e$ the elementary charge. We assume a sinusoidal modulation $V(x) = V_0 \cos (2 \pi x /a)$ throughout the paper.
The magnetic field $B$ is applied perpendicular ($\parallel z$ axis) to the 2DEG plane ($x$-$y$ plane, see Fig.\ \ref{sampleschem}).
Oscillations were also observed in the transverse direction $R_{yy}$, albeit with much smaller amplitudes and taking maxima instead of minima at Eq.\ (\ref{flatband}) \cite{Weiss89}. Soon after the discovery, a pictorial explanation invoking the $\boldsymbol{E}\times\boldsymbol{B}$ drift velocity of semiclassical cyclotron orbits was presented \cite{Beenakker89}, which captures the physics behind the dominant mechanism (\textit{band contribution}, ascribed to the modulation of the Landau-band dispersion and hence of the group velocity) generating the oscillations in $R_{xx}$. However, full understanding of CO in a ULSL, including the oscillations in $R_{yy}$, requires \cite{Gerhardts90C} quantum mechanical theories  \cite{Vasilopoulos89,Gerhardts89,Zhang90,Peeters92}, in which additional contribution due to the modulation of the density of states (\textit{collisional contribution}) is implemented. 

Although occasionally overlooked, the theories \cite{Vasilopoulos89,Gerhardts89,Zhang90,Peeters92} predict the presence of CO also in the Hall resistance $R_{yx}$, resulting from the collisional contribution as is the case in $R_{yy}$. To the knowledge of the present authors, however, an unambiguous experimental observation of CO in the Hall resistance of a ULSL has never been reported for nearly three decades after the theoretical predictions
\footnote{By contrast, CO in the Hall resistance was reported for two-dimensional antidot square superlattices as early as in 1996 \cite{Tsukagoshi96H}}.
We surmise that the observation has been hampered mainly by two obstacles: the smallness of the amplitudes and unintentional mixing of the $R_{xx}$ component into the measurement. First, the amplitude of the oscillatory part $\delta R_{yx}$ is predicted to be of the order of 1 $\Omega$, accounting for only $\sim$0.1\% of the total $R_{yx}$ $\agt$ 1 k$\Omega$ $\gg$ $R_{xx}$. The signal from $\delta R_{yx}$ can thus readily be buried in the noise level for the measurement setup with sensitivity adjusted to measure $R_{yx}$. Second, due to inevitable imperfectness of the Hall bar device, e.g., the misalignment of voltage probes, a small portion of $R_{xx}$ can inadvertently mix into the measured $R_{yx}$. The effect of the mixed $R_{xx}$ is totally insignificant in the usual measurement of $R_{yx}$, since $R_{xx}$$\ll$$|R_{yx}|$ for $B \agt 0.1$ T in high-mobility 2DEGs. Focusing on the oscillatory parts, however, parasitic $\delta R_{xx}$ component can easily outweigh the intrinsic $\delta R_{yx}$, since the amplitudes of the former is about two orders of magnitude larger than the predicted amplitudes of the latter. Here and in what follows, we denote the oscillatory part of a quantity $X$ by $\delta X$, and the difference in $X$ with and without $V(x)$ by $\Delta X$. The latter can contain the nonoscillatory part induced by $V(x)$ in addition to $\delta X$.

In the present study, we circumvent these problems by employing simple techniques: directly measuring the excess Hall resistivity $\Delta R_{yx}$ attributable to $V(x)$ and then extracting the antisymmetric part with respect to the magnetic field. The $\delta R_{yx}$ thus obtained is compared with $\delta R_{yx}$ numerically calculated from the formula for the conductivity $\sigma_{yx}$ expressed in terms of summation over the Landau indices given in Ref.\ \cite{Peeters92}. To gain transparent insight into the behavior of $\sigma_{yx}$ and to efficiently extract the oscillatory part, we also deduce an analytic asymptotic expression that approximates the $\sigma_{yx}$ quite well. We find that the observed $\delta R_{yx}$ is much smaller than the theoretical prediction, even if we consider damping of the oscillations due to small angle scatterings neglected in the original theory. 

The present study is partly motivated by rather counterintuitive isotropic behavior of the CO in the Seebeck coefficients (diagonal components of the thermopower tensor) of a ULSL predicted in Ref.\ \cite{Peeters92}, which we have recently noticed \cite{Koike20TPCO} to be strongly related via the Mott relation to the CO in the Hall conductivity. As we will see, the smallness of $R_{yx}$ found in the present study casts doubt on the isotropic behavior.

\section{Experimental details and results\label{SecExp}}

Figure \ref{sampleschem} illustrates the schematics of the Hall bar device used in the present study. The device contains a modulated area (ULSL) and a plain 2DEG (p2DEG) area in series, with the voltage probes to measure the magnetoresistance $R_{xx,\mathrm{U / P}}$ and the Hall resistance $R_{yx,\mathrm{U / P}}$ attached to both areas, where the subscript U and P represent ULSL and p2DEG areas, respectively. The device was fabricated from a conventional GaAs/AlGaAs 2DEG wafer having the mobility $\mu = 70$ m$^2$/(Vs) and the electron density $n_e = 2.1\times10^{15}$ m$^{-2}$. Modulation $V(x)$ with the period $a = 184$ nm was introduced by placing a grating of negative-tone electron-beam resist on the surface of the ULSL area \cite{Endo00e}, exploiting the strain-induced piezoelectric effect \cite{Endo00e,Skuras97}. All the measurements in this study were performed at 4.2 K\@.

\begin{figure}
\includegraphics[width=8.6cm,clip]{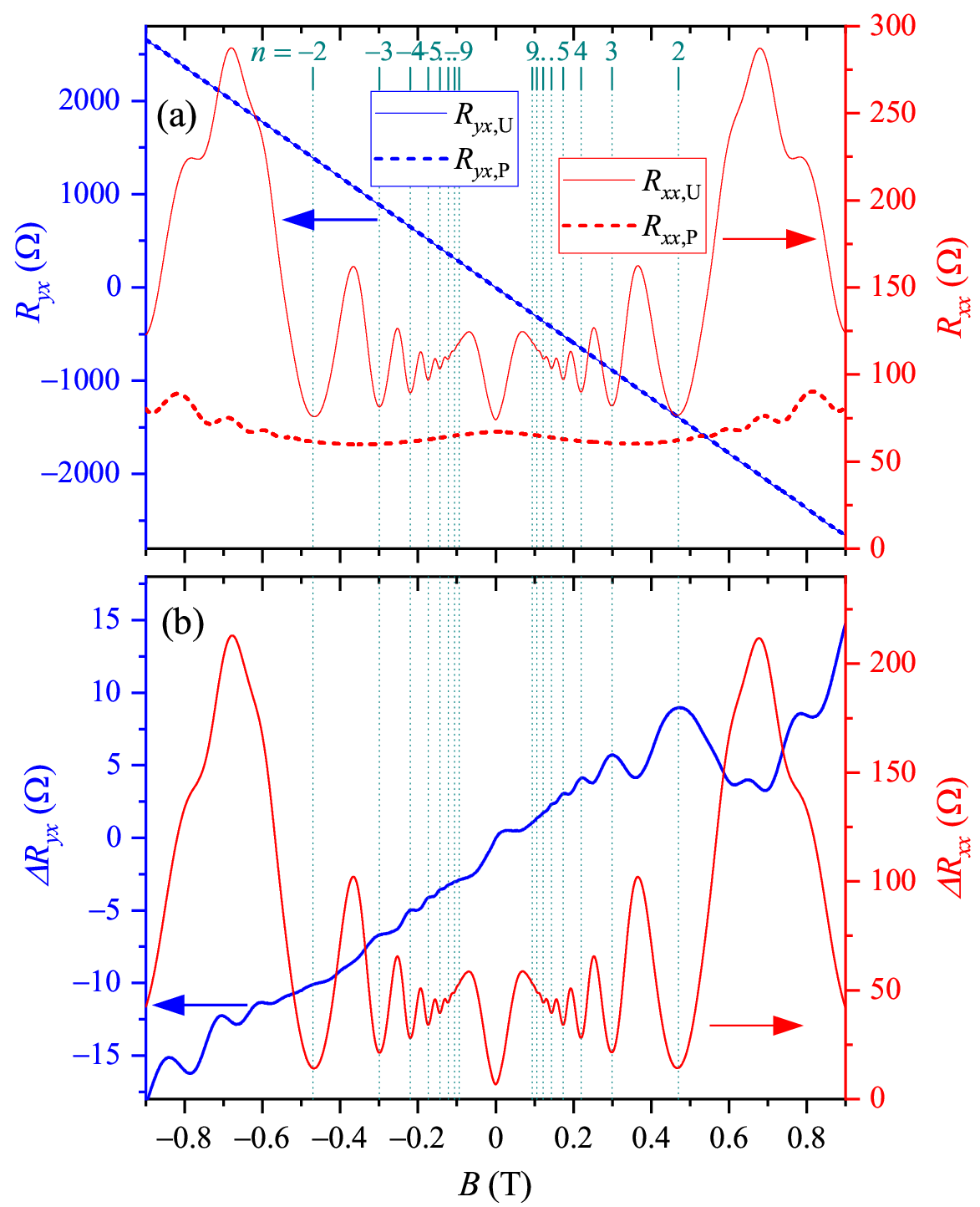}%
\caption{(a) Hall resistances $R_{yx,\mathrm{U}}$ and $R_{yx,\mathrm{P}}$ (blue lines, left axis) and magnetoresistances  $R_{xx,\mathrm{U}}$ and $R_{xx,\mathrm{P}}$ (red lines, right axis) measured in the ULSL area (solid lines) and in the p2DEG area (dashed lines).  (b) Excess Hall resistance $\Delta R_{yx}$ obtained by the arrangement depicted in Fig.\ \ref{sampleschem} (blue line, left axis) and excess magnetoresistance $\Delta R_{xx}$ obtained by taking the difference between  $R_{xx,\mathrm{U}}$ and $R_{xx,\mathrm{P}}$ shown in (a). Vertical dotted lines indicate the locations of the $n$-th flat-band condition given by Eq.\ (\ref{flatband}).\label{RandDR}}
\end{figure}

In Fig.\ \ref{RandDR}(a), we plot $R_{yx,\mathrm{U / P}}$ and $R_{xx,\mathrm{U / P}}$ measured employing standard low-frequency ($f =73$ Hz) ac lock-in technique with the current $I = 100$ nA. 
$R_{xx,\mathrm{U}}$ exhibits prominent CO with the minima occurring at the positions given by Eq.\ (\ref{flatband}). 
Small-amplitude oscillations observed at higher magnetic-field regions ($|B| \agt 0.5$ T) both in $R_{xx,\mathrm{U}}$ and $R_{xx,\mathrm{P}}$ are the Shubnikov-de Haas (SdH) oscillations.
On the other hand, $R_{yx,\mathrm{U / P}}$ appears as a featureless line in the plots. 
To extract the component deriving from $V(x)$, we take the differences $\Delta R_{yx} = R_{yx,\mathrm{U}} - R_{yx,\mathrm{P}}$ and $\Delta R_{xx} = R_{xx,\mathrm{U}} - R_{xx,\mathrm{P}}$, and plot them in Fig.\ \ref{RandDR}(b). Since the difference is large for $R_{xx}$, $\Delta R_{xx}$ can be obtained reliably by simply subtracting the two traces in Fig.\ \ref{RandDR}(a) numerically. We can see that the SdH oscillations are partially canceled out in $\Delta R_{xx}$  \footnote{Imperfectness of the cancellation are mainly attributable to the modulation of the SdH amplitude by $V(x)$. See., e.g., \cite{Endo08ModSdH}}.
In $R_{yx}$, by contrast, minuscule difference ($\sim$ $\Omega$) unobservable in Fig.\ \ref{RandDR}(a) needs to be drawn out from orders of magnitude larger ($\sim$ k$\Omega$) values. To do this with sufficient signal-to-noise (S/N) ratio, we collect the excess Hall resistance $\Delta R_{yx}$ directly, employing the arrangement depicted in Fig.\ \ref{sampleschem}: The Hall voltages from ULSL and p2DEG areas are first amplified ($\times$100) by separate differential preamplifiers \footnote{LI-75A, NF Corporation}, and then their outputs are plugged into differential input of a lock-in amplifier  \footnote{LI 5660, NF Corporation}. The input voltage range of the lock-in amplifier can thus be adjusted to the minimum range that encompasses the small difference voltage, which serves to significantly improve the S/N ratio.
As can be seen in Fig.\ \ref{RandDR}(b), $\Delta R_{yx}$ obtained by this method clearly shows oscillations corresponding to both CO and partially canceled SdH oscillations (or, more precisely, incipient quantum Hall plateaus). In the present paper, we focus on the CO\@. A notable feature is the asymmetry between $B>0$ and $B<0$ regions. In both regions, maxima are observed roughly at the flat-band conditions Eq.\ (\ref{flatband}). However, the amplitudes of the oscillations are much larger in $B>0$. As mentioned earlier, the observed CO are considered to be composed of two components: intrinsic $\delta R_{yx}$ and parasitic $\delta R_{xx}$. Since $\delta R_{yx}$ is an odd function of $B$ while $\delta R_{xx}$ is an even function, the two components are superposed either destructively ($B<0$ in the present case) or constructively ($B>0$), depending on the sign of the magnetic field. This explains the observed asymmetry in the CO amplitudes. We have measured several ULSL devices in addition to the one shown in Fig.\ \ref{RandDR}. Similar asymmetry was observed for all of them.
 
 \begin{figure}
\includegraphics[width=8.6cm,clip]{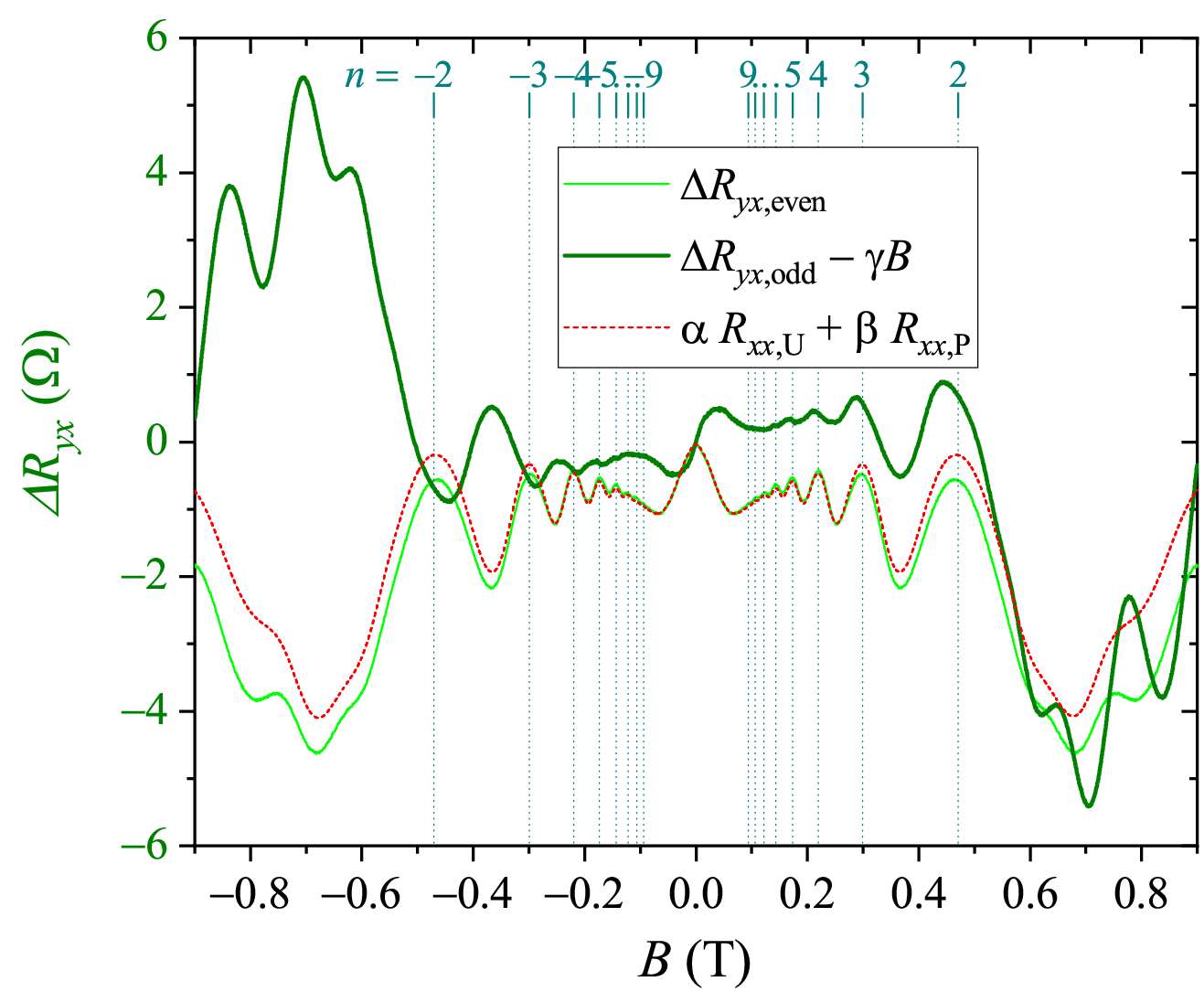}%
\caption{Even ($\Delta R_{yx,\mathrm{even}}$, thin solid line) and odd ($\Delta R_{yx,\mathrm{odd}}$, thick solid line) parts of the excess Hall resistance $\Delta R_{yx}$. Linear component $\gamma B$ (with $\gamma = 18.8$ $\Omega$/T) is subtracted from $\Delta R_{yx,\mathrm{odd}}$ for clarity. Linear combination of $R_{xx,\mathrm{U}}$ and $R_{xx,\mathrm{P}}$ (with $\alpha = -0.0197$ and $\beta = 0.0212$), which emulates the parasitic $R_{xx}$ component contained in the measured Hall resistance, is also shown (thin dashed line).  Vertical dotted lines indicate the locations of the $n$-th flat-band condition given by Eq.\ (\ref{flatband}).\label{SymASymPM}}
\end{figure}

\begin{figure}
\includegraphics[width=8.6cm,clip]{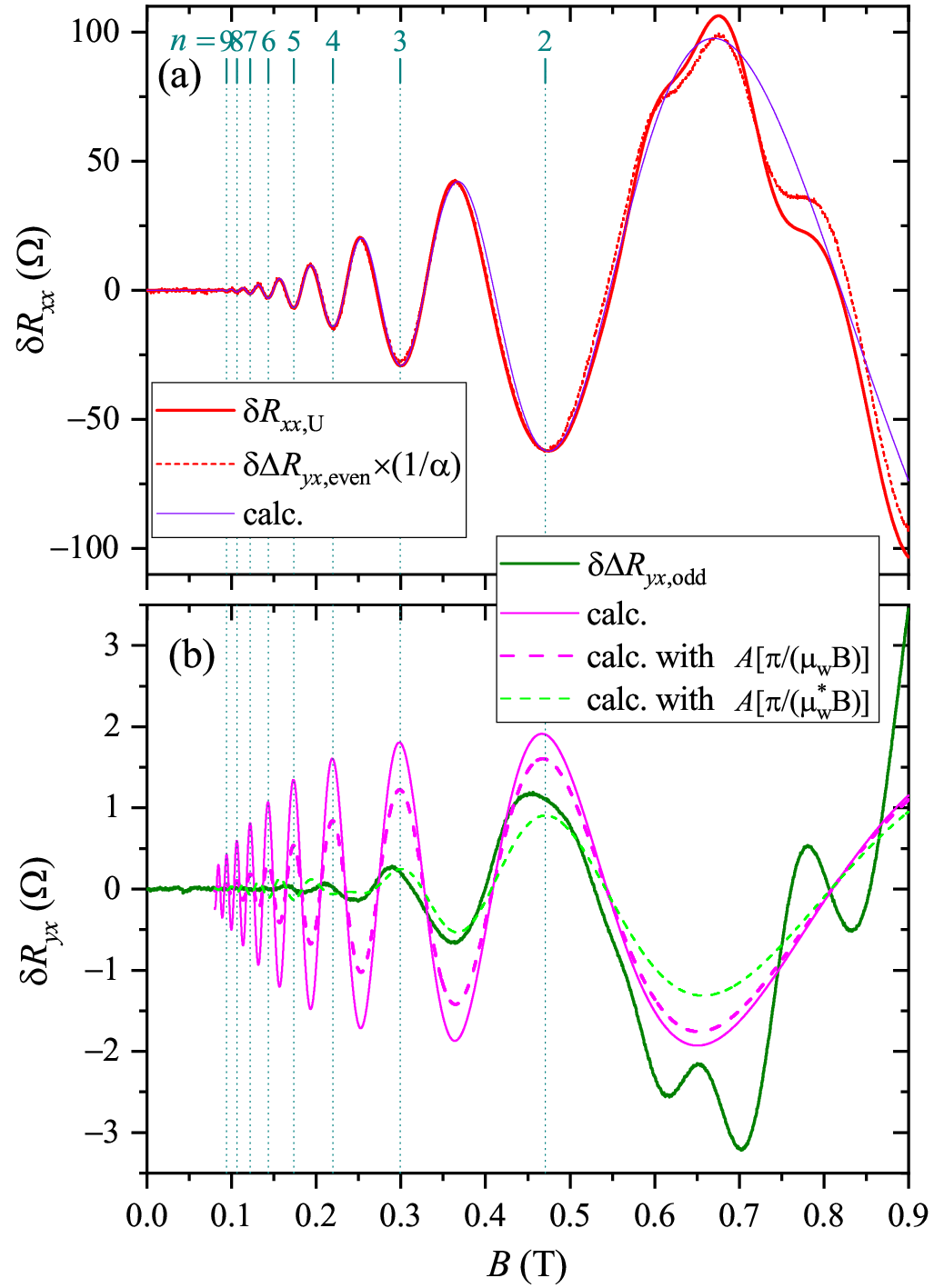}%
\caption{(a) Oscillatory parts $\delta R_{xx,\mathrm{U}}$ (thick solid line) and $\delta \Delta R_{yx,\mathrm{even}}$ (thin dashed line) obtained by subtracting the slowly-varying background from $R_{xx,\mathrm{U}}$ and $\Delta R_{yx,\mathrm{even}}$ shown in Figs.\ \ref{RandDR}(a) and \ref{SymASymPM}, respectively. The latter is scaled by the factor $1/\alpha = -50.6$. Calculated $\delta R_{xx}$ [Eq.\ (\ref{deltaRxx}) with $V_0 = 0.35$ meV and $\mu_\mathrm{w} = 8.2$ m$^2$/(Vs)] is also shown (thin solid line), exhibiting excellent agreement with $\delta R_{xx,\mathrm{U}}$ (apart from SdH oscillations at $B \agt 0.6$ T). (b) Oscillatory parts $\delta \Delta R_{yx,\mathrm{odd}}$ (thick solid line), extracted from the experimentally obtained $\Delta R_{yx,\mathrm{odd}}$ shown in Fig.\ \ref{SymASymPM} by subtracting the slowly-varying background. $\delta R_{yx}$ calculated with Eq.\ (\ref{deltaRyx}) (thin solid line) and modified Eq.\ (\ref{deltaRyx}) with the damping factor $A[\pi/(\mu_\mathrm{w} B)]$ multiplied to $\delta \sigma_{yx}^\mathrm{A}$ (dashed line), using the sample parameters deduced in (a), are also plotted. Thin dashed curve represents basically the same calculation but with $\mu_\mathrm{w}$ in the damping factor replaced by $\mu_\mathrm{w}^* = \mu_\mathrm{w}/2$. Vertical dotted lines indicate the locations of the $n$-th flat-band condition given by Eq.\ (\ref{flatband}).\label{oscparts}}
\end{figure}

In order to separate the two components, we take the even $\Delta R_{yx,\mathrm{even}}$ and the odd  $\Delta R_{yx,\mathrm{odd}}$ parts of $\Delta R_{yx}$, $\Delta R_{yx,\mathrm{even}}(B) = [\Delta R_{yx}(B) + \Delta R_{yx}(-B)]/2$ and $\Delta R_{yx,\mathrm{odd}}(B) = [\Delta R_{yx}(B) - \Delta R_{yx}(-B)]/2$, corresponding to the parasitic and the intrinsic components, respectively, and plot them in Fig.\ \ref{SymASymPM}. Noting that the Hall probes in both ULSL and p2DEG areas generally can pick up the corresponding parasitic $R_{xx,\mathrm{U / P}}$ components independently, possibly with differing weights, we can expect that the parasitic component can be expressed by the linear combination $\alpha R_{xx,\mathrm{U}} + \beta R_{xx,\mathrm{P}}$ with small values of $|\alpha|$ and $|\beta|$. By properly selecting $\alpha$ and $\beta$, with special care to reproduce the oscillatory part due to the CO [see also Fig.\ \ref{oscparts}(a)], fairly good agreement can be achieved between $\alpha R_{xx,\mathrm{U}} + \beta R_{xx,\mathrm{P}}$ and the observed $\Delta R_{yx,\mathrm{even}}$, supporting the interpretation on the origin of $\Delta R_{yx,\mathrm{even}}$. This confirms that the remnant $\Delta R_{yx,\mathrm{odd}}$ is the intrinsic CO in $R_{yx}$, the target we are seeking in the present study. In the plot of the odd part, we subtracted a linear term $\gamma B$, which is attributable to the small difference in the electron density, $\Delta n_e \simeq - 1.3\times10^{13}$ m$^{-2}$, between the ULSL and the p2DEG areas. 
 
\section{Comparison with Theoretical Calculations\label{SecComp}}
\subsection{Deducing superlattice parameters from commensurability oscillations in the magnetoresistance\label{SubSecRxx}}

The next step is to compare the observed $\delta R_{yx}$ with the theoretical prediction. 
Before discussing $\delta R_{yx}$,  however, we briefly review well-documented behavior of $\delta R_{xx}$ \cite{Peeters92,Endo00e}, from which we draw out parameters characterizing our ULSL\@. As mentioned earlier, two different mechanisms, the band and the collisional contributions, are responsible for CO\@. 
Asymptotic analytic expressions for the oscillatory parts of the conductivity, valid at low magnetic fields where large numbers of Landau levels are occupied \cite{Peeters92}, are given for the two contributions as,
\begin{equation}
\delta \sigma_{yy}^\mathrm{band} = \frac{\sigma_0 V_0^2}{E_\mathrm{F} \hbar \omega_\mathrm{c} a k_\mathrm{F}} A\left( \frac{\pi}{\mu_\mathrm{w} B} \right) A\left( \frac{T}{T_a} \right) \sin{r_\mathrm{c}} \label{Dsgmband}
\end{equation}
and
\begin{equation}
\delta \sigma_{xx}^\mathrm{col} = -\frac{3 \sigma_0 V_0^2 a k_\mathrm{F}}{8 \pi^2 E_\mathrm{F} \hbar \omega_\mathrm{c} (\mu B)^2} A\left( \frac{\pi}{\mu_\mathrm{w} B} \right) A\left( \frac{T}{T_a} \right) \sin{r_\mathrm{c}}, \label{Dsgmcol}
\end{equation}
respectively, where $\sigma_0 = e n_e \mu$ is the conductivity at $B = 0$, $E_\mathrm{F}$ the Fermi energy, $\omega_\mathrm{c} = e |B|/m^*$ the cyclotron angular frequency with $m^*$ the effective mass, $r_\mathrm{c} \equiv 4\pi R_\mathrm{c} / a$, $T_a \equiv \hbar \omega_\mathrm{c} a k_\mathrm{F} / (4 \pi^2 k_\mathrm{B})$, and $A(x) \equiv x/\sinh x$. Although absent in the original theories \cite{Vasilopoulos89,Gerhardts89,Zhang90,Peeters92}, an additional damping factor $A[\pi/(\mu_\mathrm{w} B)]$ accounting for the effect of small angle scattering, with the value of $\mu_\mathrm{w}$ close to the quantum mobility $\mu_\mathrm{q}$ \cite{Mirlin98,Endo00e},  are contained in Eqs.\ (\ref{Dsgmband}) and (\ref{Dsgmcol}) in addition to the thermal damping factor $A(T/T_a)$.
More detailed discussion on the factor $A[\pi/(\mu_\mathrm{w} B)]$ will be given below.
The resistivity tensor $\rho_{ij}$ ($i,j = x,y$) is obtained by inverting the conductivity tensor $\sigma_{ij}$: $\sigma_{xx} = \sigma_{xx}^\mathrm{sc}+\delta \sigma_{xx}^\mathrm{col}$,  $\sigma_{yy} = \sigma_{xx}^\mathrm{sc}+\delta \sigma_{xx}^\mathrm{col}+\delta \sigma_{yy}^\mathrm{band}$, and $\sigma_{yx} = -\sigma_{xy} = \sigma_{yx}^\mathrm{sc}+\delta \sigma_{yx}$, where $\sigma_{xx}^\mathrm{sc} = \sigma_0 / (1 + \mu^2 B^2)$ and $\sigma_{yx}^\mathrm{sc} = \sigma_0 \mu B / (1 + \mu^2 B^2)$ are the semiclassical conductivities for a p2DEG\@. Noting that $\mu B \gg 1$ and $|\delta \sigma_{xx}^\mathrm{band}| \gg |\delta \sigma_{xx}^\mathrm{col}|$, and using the relation $R_{xx} / R_0 = \sigma_0 \rho_{xx}$ with $R_0$ representing $R_{xx}$ at $B = 0$, we obtain, to a good approximation, 
\begin{equation}
\frac{\delta R_{xx}}{R_0} = (\mu B)^2 \frac{\delta \sigma_{yy}^\mathrm{band}}{\sigma_0} \label{deltaRxx}
\end{equation}
and likewise $\delta R_{yy} / R_0 = (\mu B)^2 \delta \sigma_{xx}^\mathrm{col} / \sigma_0$. Equation (\ref{deltaRxx}) has been shown to describe experimentally obtained CO extremely well \cite{Endo00e}. This is confirmed in Fig.\ \ref{oscparts}(a), which shows $\delta R_{xx,\mathrm{U}}$ extracted from $R_{xx,\mathrm{U}}$ in Fig.\ \ref{RandDR}(a) by subtracting slowly varying background following the protocol detailed in Ref.\ \cite{Endo00e}, along with $\delta R_{xx}$ in Eq.\ (\ref{deltaRxx}) obtained by the fitting, employing $V_0$ and $\mu_\mathrm{w}$ as fitting parameters. The fitting yields $V_0 = 0.35$ meV and $\mu_\mathrm{w} = 8.2$ m$^2$/(Vs). 
The value of $\mu_\mathrm{w}$ is close to $\mu_\mathrm{q} = 8.6$  m$^2$/(Vs) deduced from the SdH oscillations in $R_{xx,\mathrm{P}}$ plotted in Fig.\ \ref{RandDR}(a).

\begin{figure}[b]
\includegraphics[width=8.6cm,clip]{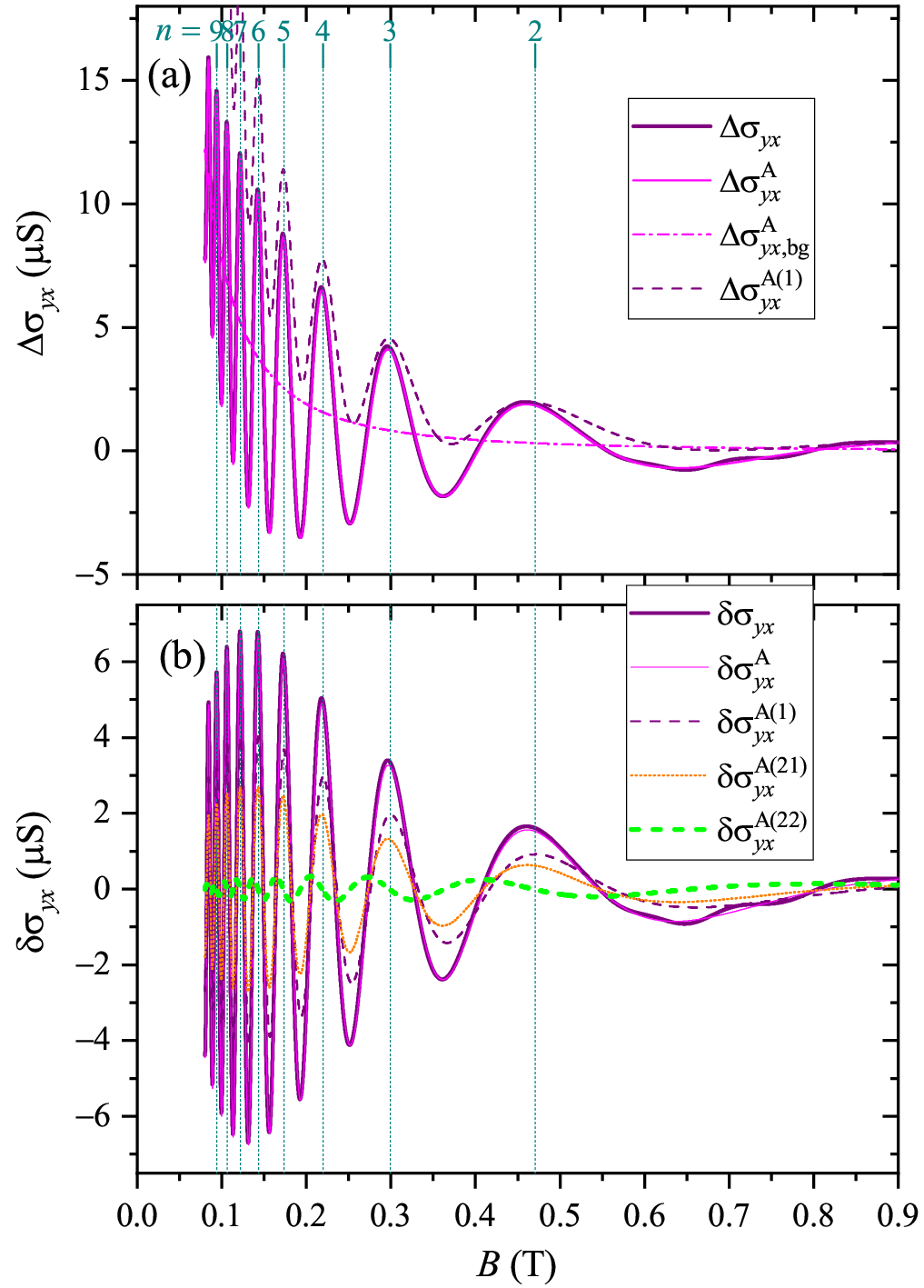}%
\caption{(a) The increment of the Hall conductivity due to $V(x)$, calculated with the exact [thick solid line, $\Delta \sigma_{yx}$ in Eq.\ (\ref{Dsgmyx})] and the approximate analytic [thin solid line, $\Delta \sigma_{yx}^{\mathrm{A}}$ in Eq.\ (\ref{DsgmyxA})] formulas, using sample parameters of the ULSL in the present study. Non-oscillatory background, $\Delta \sigma_{yx,\mathrm{bg}}^\mathrm{A}$ in Eq.\ (\ref{DsgmyxAbg}), of the approximate formula is plotted with thin dot-dashed line. The increment deriving only from the first term in Eq.\ (\ref{sgmyxA}), namely, $\Delta \sigma_{yx}^{\mathrm{A}(1)} =  \sigma_{yx,\mathrm{bg}}^{\mathrm{A}(1)} + \delta \sigma_{yx}^{\mathrm{A}(1)}$ in Eq.\ (\ref{sgmyxA1}), is also plotted (dashed line). (b) Oscillatory parts calculated from exact (thick solid line, $\delta \sigma_{yx} = \Delta \sigma_{yx}-\Delta \sigma_{yx,\mathrm{bg}
}^\mathrm{A}$) and approximate [thin solid line, $\delta \sigma_{yx}^\mathrm{A}$ in Eq.\ (\ref{dsgmyxA})] formulas. Three constituent terms of the oscillatory part are also plotted separately, with thin dashed, thin dotted, and thick dashed lines representing $\delta \sigma_{yx}^{\mathrm{A}(1)}$, $\delta \sigma_{yx}^{\mathrm{A}(21)}$, and $\delta \sigma_{yx}^{\mathrm{A}(22)}$ in Eqs.\ (\ref{dsgmyxA1}), (\ref{dsgmyxA21}), and (\ref{dsgmyxA22}), respectively. Vertical dotted lines indicate the locations of the $n$-th flat-band condition given by Eq.\ (\ref{flatband}).\label{sgmyxcalc}}
\end{figure}

\subsection{Asymptotic analytic expressions for the Hall conductivity\label{SubSecAsympt}}

Now we turn to the Hall component. We start with the expression of the Hall conductivity in a ULSL presented in Ref.\ \cite{Peeters92},
\begin{equation}
\sigma_{yx} = \frac{2e^2}{h}\ \sum_{N=0}^{\infty} {(N+1) \int_{0}^{1}{\frac{f_{E_\mathrm{F}}\left(E_{N,\xi}\right)-f_{E_\mathrm{F}}\left(E_{N+1,\xi}\right)}{\left[1+\lambda_N\cos{\left(2\pi\xi\right)}\right]^2}} d\xi\ } \label{sgmyxnSum}
\end{equation}
with
\begin{equation}
E_{N,\xi} = E_N+V_0 \exp \left(-\frac{u}{2}\right) L_N\left(u\right) \cos{\left(2\pi\xi\right)} \label{Enxi}
\end{equation}
and
\begin{equation}
\lambda_N = \frac{V_0}{\hbar \omega_\mathrm{c}} \exp\left( -\frac{u}{2} \right) L_{N+1}^{-1}(u),  \label{lambdaNPV}
\end{equation}
where $f_{E_\mathrm{F}}(E) =\{1+\exp[(E-E_\mathrm{F})/k_\mathrm{B}T]\}^{-1}$ is the Fermi-Dirac distribution function, $N$ is the Landau index,  $E_N \equiv (N+1/2)\hbar \omega_\mathrm{c}$, $\xi \equiv x_0/a$ with $x_0$ the guiding center, $u \equiv 2 \pi^2 l^2 / a^2$ with $l = \sqrt{\hbar / (e|B|)}$ the magnetic length, and $L_N(u)$ and $L_{N}^{M}(u)$ are the Laguerre and the associated Laguerre polynomials. Note that Eq.\ (\ref{sgmyxnSum}) is valid only for $B > 0$. Since $\sigma_{yx}$ is an antisymmetric function with respect to $B$, $\sigma_{yx}$ at $B<0$ is obtained by inverting the sign of Eq.\ (\ref{sgmyxnSum}). %The same relation holds for all the approximate formulas derived from Eq.\ (\ref{sgmyxnSum}) presented below. 
The increment of $\sigma_{yx}$ introduced by the modulation, 
\begin{equation}
\Delta \sigma_{yx} = \sigma_{yx}(V_0) - \sigma_{yx}(V_0 = 0), \label{Dsgmyx}
\end{equation}
numerically calculated 
\footnote{The upper limit of the summation was truncated  at $N = 70$, which is large enough for the temperature and magnetic-field range considered} using Eq.\ (\ref{sgmyxnSum}) with the parameters in the present ULSL is plotted in Fig.\ \ref{sgmyxcalc}(a). 

Since the behavior of $\sigma_{yx}$, notably the phase of the oscillations, are not readily perceived from Eq.\ (\ref{sgmyxnSum}), we deduce an asymptotic analytic expression, basically following the prescription taken for $\sigma_{xx}$ and $\sigma_{yy}$ in Ref.\ \cite{Peeters92}. Via the deriving procedure detailed in the Appendix, we arrive at an approximate formula $\sigma_{yx}^\mathrm{A} \simeq \sigma_{yx}$, 
\begin{equation}
\sigma_{yx}^\mathrm{A} = \sigma_{yx}^{\mathrm{A}(1)} + \sigma_{yx}^{\mathrm{A}(21)}+ \sigma_{yx}^{\mathrm{A}(22)}, \label{sgmyxA} %= \nu\frac{e^2}{h} + \Delta \sigma_{yx}^\mathrm{A}
\end{equation}
with
\begin{subequations}
\begin{align}
\sigma_{yx}^{\mathrm{A}(1)} & = \nu\frac{e^2}{h} + \Delta \sigma_{yx}^{\mathrm{A}(1)} = \nu\frac{e^2}{h} + \sigma_{yx,\mathrm{bg}}^{\mathrm{A}(1)} + \delta \sigma_{yx}^{\mathrm{A}(1)} \label{sgmyxA1} \\
%\begin{equation}
& \sigma_{yx,\mathrm{bg}}^{\mathrm{A}(1)} =  \nu\frac{e^2}{h} \frac{3}{2}{\lambda_\mathrm{c}}^2 \label{sgmyxA1bg} \\
%\end{equation}
%\begin{equation}
& \delta \sigma_{yx}^{\mathrm{A}(1)} =  -\nu\frac{e^2}{h}A\left(\frac{T}{T_{a}}\right)\frac{3}{2}{\lambda_\mathrm{c}}^2\sin{\left(r_\mathrm{c}+\delta_\mathrm{F}\right)}, \label{dsgmyxA1}
%\end{equation}
\end{align}
\end{subequations}
\begin{widetext}
\begin{subequations}
\begin{align}
\sigma_{yx}^{\mathrm{A}(21)} & =  \sigma_{yx,\mathrm{bg}}^{\mathrm{A}(21)} + \delta \sigma_{yx}^{\mathrm{A}(21)} \label{sgmyxA21} \\
& \sigma_{yx,\mathrm{bg}}^{\mathrm{A}(21)} =  -\nu\frac{e^2}{h}\frac{ak_\mathrm{F}}{\pi}\sin{\left(\frac{\pi}{{ak}_\mathrm{F}}\right)}{\lambda_\mathrm{c}}^2\cos{\left(\frac{\delta_\mathrm{F}}{2}+\frac{\pi}{ak_\mathrm{F}}\right)} \label{sgmyxA21bg}  \\
%\begin{equation}
& \delta \sigma_{yx}^{\mathrm{A}(21)} = -\nu\frac{e^2}{h}\frac{ak_\mathrm{F}}{\pi}\sin{\left(\frac{\pi}{{ak}_\mathrm{F}}\right)}A\left(\frac{T}{T_a}\right){\lambda_\mathrm{c}}^2\sin{\left(r_\mathrm{c}+\frac{\delta_\mathrm{F}}{2}-\frac{\pi}{{ak}_\mathrm{F}}\right)}, \label{dsgmyxA21} 
%\end{equation}
\end{align}
\end{subequations}
and
\begin{subequations}
\begin{align}
\sigma_{yx}^{\mathrm{A}(22)} & = \sigma_{yx,\mathrm{bg}}^{\mathrm{A}(22)} + \delta \sigma_{yx}^{\mathrm{A}(22)} \label{sgmyxA22}  \\
& \sigma_{yx,\mathrm{bg}}^{\mathrm{A}(22)}  =  - \mathrm{sgn}(B) \frac{e^2}{h}\frac{ak_\mathrm{F}}{\pi}\cos{\left(\frac{\pi}{{ak}_\mathrm{F}}\right)}{\lambda_\mathrm{c}}^2\sin{\left(\frac{\delta_\mathrm{F}}{2}+\frac{\pi}{{ak}_\mathrm{F}}\right)} \label{sgmyxA22bg} \\
& \delta \sigma_{yx}^{\mathrm{A}(22)} =  \mathrm{sgn}(B) \frac{e^2}{h}\frac{ak_\mathrm{F}}{\pi}\cos{\left(\frac{\pi}{{ak}_\mathrm{F}}\right)}A\left(\frac{T}{T_a}\right){\lambda_\mathrm{c}}^2\cos{\left(r_\mathrm{c}+\frac{\delta_\mathrm{F}}{2}-\frac{\pi}{{ak}_\mathrm{F}}\right)}, \label{dsgmyxA22}
\end{align}
\end{subequations}
\end{widetext}
%\begin{widetext}
%\begin{subequations}
%\begin{align}
%\delta \sigma_{yx}^{A(21)} = & \left(\nu+1\right)\frac{e^2}{h}\sqrt{\frac{2}{\pi}}\frac{V_0}{\hbar\omega_c}\frac{\lambda_0}{\sqrt{r_\mathrm{c}^-}}\ \left[A\left(\frac{T}{T_a}\right)\cos{\left(\frac{r_\mathrm{c}^-+r_\mathrm{c}^+}{2}+\frac{\delta_\mathrm{F}^+}{2}\right)}-\sin{\left(\frac{\delta_\mathrm{F}^+}{2}-\frac{r_\mathrm{c}^--r_\mathrm{c}^+}{2}\right)}\right] \label{dsgmyxA21} \\
%\end{equation}
%\begin{equation}
%\delta \sigma_{yx}^{A(22)} = & \left(\nu-1\right)\frac{e^2}{h}\sqrt{\frac{2}{\pi}}\frac{V_0}{\hbar\omega_c}\frac{\lambda_0}{\sqrt{r_\mathrm{c}}}\left[A\left(\frac{T}{T_a}\right)\cos{\left(r_\mathrm{c}^-+\frac{\delta_\mathrm{F}^-}{2}\right)}-\frac{1}{\sqrt{1+{r_\mathrm{c}}^2}}\right], \label{dsgmyxA22}
%\end{align}
%\end{subequations}
%\end{widetext}
where $\nu = n_e h / (eB)$ is the Landau-level filling factor ($\nu < 0$ for $B < 0$ by the definition), $\lambda_\mathrm{c} \equiv 2\sqrt{2/\pi} [V_0/ (\hbar\omega_c)] [\pi/(ak_\mathrm{F})]{r_\mathrm{c}}^{-1/2}$, $\delta_\mathrm{F} \equiv 2 \cot^{-1}{r_\mathrm{c}}$, and $\mathrm{sgn}(x)$ represents the sign of $x$.
By collecting the corresponding terms,  we obtain the increment of the conductivity, the nonoscillatory background of the increment, and the oscillatory part,
\begin{equation}
\Delta \sigma_{yx}^\mathrm{A} = \Delta \sigma_{yx}^{\mathrm{A}(1)} + \sigma_{yx}^{\mathrm{A}(21)}+ \sigma_{yx}^{\mathrm{A}(22)}, \label{DsgmyxA} %- \nu\frac{e^2}{h}
\end{equation}
\begin{equation}
\Delta \sigma_{yx,\mathrm{bg}}^\mathrm{A} = \sigma_{yx,\mathrm{bg}}^{\mathrm{A}(1)} + \sigma_{yx,\mathrm{bg}}^{\mathrm{A}(21)}+ \sigma_{yx,\mathrm{bg}}^{\mathrm{A}(22)}, \label{DsgmyxAbg}
\end{equation}
and
\begin{equation}
\delta \sigma_{yx}^\mathrm{A} = \delta \sigma_{yx}^{\mathrm{A}(1)} + \delta \sigma_{yx}^{\mathrm{A}(21)}+ \delta \sigma_{yx}^{\mathrm{A}(22)}, \label{dsgmyxA}
\end{equation}
respectively. Figure \ref{sgmyxcalc}(a) illustrates that the asymptotic analytic expression Eq.\ (\ref{DsgmyxA}) reproduces Eq.\ (\ref{Dsgmyx}) quite well, except for the small-amplitude oscillations at $B \agt 0.6$ T resulting from the Landau quantization. This allows us to use the approximate background $\Delta \sigma_{yx,\mathrm{bg}}^\mathrm{A}$ to extract the oscillatory part from $\Delta \sigma_{yx}$ in Eq.\ (\ref{Dsgmyx}). The oscillatory part thus obtained, $\delta \sigma_{yx} = \Delta \sigma_{yx} - \Delta \sigma_{yx,\mathrm{bg}}^\mathrm{A}$, is plotted in Fig.\ \ref{sgmyxcalc}(b) along with $\delta \sigma_{yx}^\mathrm{A}$ in Eq.\ (\ref{dsgmyxA}). 

The analytic expression lets us grasp the outline of the behavior of the oscillations. Since $\delta \sigma_{yx}^{\mathrm{A}(1)}$, $\delta \sigma_{yx}^{\mathrm{A}(21)}$, and $\delta \sigma_{yx}^{\mathrm{A}(22)}$ all oscillate with different phases depending on $B$ through $\delta_\mathrm{F}$, the phase of the CO in $\sigma_{yx}$ is expected to exhibit rather complicated behavior. We note, however, that $\pi/(a k_\mathrm{F})$ ($=$0.148 in the present sample) is generally small for experimentally achievable values of $a$ and that $\delta_\mathrm{F}$ is also small ($\alt 0.4$ in the magnetic-field range where CO is observed) and approaches 0 with decreasing $B$. Furthermore, at low magnetic fields where $\nu$ is large, $\delta \sigma_{yx}^{\mathrm{A}(22)}$ becomes much smaller than the other terms. The dominance of $\delta \sigma_{yx}^{\mathrm{A}(1)}$ and $\delta \sigma_{yx}^{\mathrm{A}(21)}$, combined with the smallness of $\pi/(a k_\mathrm{F})$ and $\delta_\mathrm{F}$, indicates that the oscillation phase of $\delta \sigma_{yx}$ is close to that of $\delta \sigma_{xx}^\mathrm{col}$ and thus takes maximum at the flat-band conditions Eq.\ (\ref{flatband}), or equivalently, at $r_\mathrm{c} = 2 \pi n -\pi/2$.  The calculated $\delta \sigma_{yx}$ plotted in Fig.\ \ref{sgmyxcalc}(b) are seen to actually take maxima at the flat-band conditions at low magnetic fields. With the increase of the magnetic field, slight deviation of the peak positions becomes apparent, mainly due to the increase of $\delta_\mathrm{F}$ and of the relative importance of the third term $\delta \sigma_{yx}^{\mathrm{A}(22)}$, whose oscillation phase differs from $\delta \sigma_{yx}^{\mathrm{A}(21)}$ by $\pi/2$. Noting that $(a k_\mathrm{F}/\pi) \sin{[\pi/(a k_\mathrm{F})]} \sim$ 1, the two dominant terms are expected to have comparable oscillation amplitudes, which can also be confirmed in Fig.\ \ref{sgmyxcalc}(b).

\subsection{Comparison between experimental and calculated commensurability oscillations\\ in the Hall resistance\label{SubSecRyx}}

We obtain the Hall resistivity $\rho_{yx}$ ($= R_{yx}$ in a 2DEG) by inverting the conductivity tensor, and %noting that $R_{yx} = \rho_{yx}$ in a 2DEG,  we 
find that the oscillatory part of the Hall resistance $R_{yx}$ is given, considering $|\delta \sigma_{yx}^\mathrm{A}| \ll |\sigma_{yx}|$, by
\begin{equation}
\delta R_{yx} = \frac{1}{{\sigma_0}^2} \{ [(\mu B)^2-1] \delta \sigma_{yx}^\mathrm{A} +\mu B (2 \delta \sigma_{xx}^\mathrm{col}+\delta \sigma_{yy}^\mathrm{band}) \}. \label{deltaRyx}
\end{equation}
The oscillations are dominated by the first term. The second term is negligibly small since $|\delta \sigma_{xx}^\mathrm{col}| \ll |\delta \sigma_{yy}^\mathrm{band}|$. The third term, having the phase roughly opposite to the first term, serves to reduce the oscillation amplitude.
 In Fig.\ \ref{oscparts}(b), we compare the experimentally obtained oscillatory part $\delta \Delta R_{yx,\mathrm{odd}}$, extracted from $\Delta R_{yx,\mathrm{odd}}$ shown in Fig.\ \ref{SymASymPM} by subtracting the slowly varying background \cite{Endo00e}, with $\delta R_{yx}$ calculated by Eq.\ (\ref{deltaRyx}) using the sample parameters deduced above from the analysis of $\delta R_{xx}$. The figure shows that the observed amplitude of the CO is much smaller than the theoretical prediction especially at lower magnetic fields, while the phase of the oscillations is roughly in agreement.

It is well known that the scattering in a GaAs/AlGaAs 2DEG is predominantly caused by remote ionized donors, for which scattering angles are generally small \cite{Coleridge91}. Although the momentum relaxation is not significant for small scattering angles, cyclotron orbits are disturbed regardless of the scattering angle and thus the CO amplitudes are severely diminished even by the small-angle scattering \cite{Mirlin98,Endo00e}. The damping of the CO is more prominent for lower magnetic fields where the circumference of the cyclotron orbit  $2 \pi R_\mathrm{c}$ becomes large.
The effect of small-angle scattering, which has not been considered thus far for $\sigma_{yx}^\mathrm{A}$, can be implemented by multiplying $A[\pi/(\mu_\mathrm{w}B)]$, following the recipe applied for $\delta \sigma_{yy}^\mathrm{band}$ and $\delta \sigma_{xx}^\mathrm{col}$ described above. 
Figure \ref{oscparts}(b) reveals, however, that the discrepancy between the amplitudes of the observed and the calculated CO is still large even with the inclusion of the effect of the small-angle scattering. 
Apparently, the theory overestimates the CO amplitudes in the Hall resistance, possibly because $\delta R_{yx}$ is much more vulnerable to the scattering compared to $\delta R_{xx}$ and thus its damping cannot be described by the factor $A[\pi/(\mu_\mathrm{w}B)]$ with the same value of $\mu_\mathrm{w}$. Note that the damping factor $A[\pi/(\mu_\mathrm{w}B)]$ with $\mu_\mathrm{w} \simeq \mu_\mathrm{q}$ is firmly established theoretically \cite{Mirlin98} and experimentally \cite{Endo00e} only for $\delta \sigma_{yy}^\mathrm{band}$ and thus may not be applicable to $\delta \sigma_{xx}^\mathrm{col}$ and  $\sigma_{yx}^\mathrm{A}$ without modification 
\footnote{We have also found that $\delta R_{yy}$ calculated assuming the same damping factor as $\delta \sigma_{yy}^\mathrm{band}$ for $\delta \sigma_{xx}^\mathrm{col}$ significantly exceeds (falls below) the experimentally observed CO amplitudes at low (high) magnetic fields \cite{Koike20TPCO}}. 
As demonstrated in Fig.\ \ref{oscparts}(b), heavier damping of  $\sigma_{yx}^\mathrm{A}$ achieved by halving the $\mu_\mathrm{w}$ roughly reproduces the experimental CO amplitudes, albeit without solid theoretical underpinnings. 

\section{Possible anisotropy in the Seebeck Coefficient\label{Discussion}}

Finally, we briefly discuss the effect of $\delta \sigma_{yx}$ on the CO of the Seebeck coefficients $S_{xx}$ and $S_{yy}$, the diagonal components of the  thermopower tensor $S_{ij}$. The theory \cite{Peeters92} predicts that $S_{xx}$ and $S_{yy}$ are almost identical and thus the Seebeck coefficient accommodates CO isotropically. $S_{ij}$ can be written as the product of the resistivity tensor $\rho_{ij}$ and the thermoelectric conductivity tensor $\varepsilon_{ij}$. The latter is related to the conductivity tensor by the Mott formula \cite{Jonson84},  $\varepsilon_{ij} = -L_0 e T (d\sigma_{ij, T = 0} / d E)|_{E = E_\mathrm{F}}$ with $L_0 = \pi^2 {k_\mathrm{B}}^2/(3e^2)$ the Lorenz number, at low temperatures
\footnote{More precisely, we should use the exact formula $\varepsilon_{ij} = (1/eT) \int dE [df_{E_\mathrm{F}} (E)/dE] (E-E_\mathrm{F}) \sigma_{ij, T = 0}$, but this does not alter the qualitative argument in this section}.
We thus have $S_{xx} = \rho_{xx} \varepsilon_{xx} - \rho_{yx} \varepsilon_{yx}$ and $S_{yy} = \rho_{yy} \varepsilon_{yy} - \rho_{yx} \varepsilon_{yx}$, where we made use of the relations $\rho_{xy} = -\rho_{yx}$ and $\varepsilon_{xy} = -\varepsilon_{yx}$. The corresponding oscillatory parts due to the CO are given, to a good approximation, by $\delta S_{xx} \simeq (\delta \varepsilon_{xx} + \mu B \delta \varepsilon_{yx})/\sigma_0$ and $\delta S_{yy} \simeq (\delta \varepsilon_{yy} + \mu B \delta \varepsilon_{yx})/\sigma_0$
\footnote{We can readily show numerically that $\delta S_{xx} \simeq \rho_{xx,\mathrm{bg}} \delta \varepsilon_{xx} - \rho_{yx,\mathrm{bg}} \delta \varepsilon_{yx} + \varepsilon_{xx,\mathrm{bg}} \delta \rho_{xx}  - \varepsilon_{yx,\mathrm{bg}} \delta \rho_{yx}  \simeq \rho_{xx,\mathrm{bg}} \delta \varepsilon_{xx} - \rho_{yx,\mathrm{bg}} \delta \varepsilon_{yx} $ and $\delta S_{yy} \simeq \rho_{xx,\mathrm{bg}} \delta \varepsilon_{yy} - \rho_{yx,\mathrm{bg}} \delta \varepsilon_{yx} + \varepsilon_{xx,\mathrm{bg}} \delta \rho_{yy}  - \varepsilon_{yx,\mathrm{bg}} \delta \rho_{yx}  \simeq \rho_{xx,\mathrm{bg}} \delta \varepsilon_{yy} - \rho_{yx,\mathrm{bg}} \delta \varepsilon_{yx} $, where the subscript ``bg'' signifies the non-oscillatory background. With $\rho_{xx,\mathrm{bg}} = 1/\sigma_0$ and $\rho_{yx,\mathrm{bg}} = -\mu B/\sigma_0$, we arrive at the approximate equations presented here.}.
Since $\mu B \gg 1$ for a high-mobility 2DEG in the magnetic field range where CO can be observed, both $\delta S_{xx}$ and $\delta S_{yy}$ are dominated by the identical second term if the magnitude of $|\delta \varepsilon_{yx}|$ is comparable to those of $|\delta \varepsilon_{xx}|$ and $|\delta \varepsilon_{yy}|$ as predicted in the theory \cite{Peeters92}, leading to the rather counterintuitive isotropic behavior $\delta S_{xx} \simeq \delta S_{yy}$. However, the small $|\delta \sigma_{yx}|$ we have experimentally found in the present study, combined with the Mott formula, implies that $|\delta \varepsilon_{yx}|$ is much smaller than the theoretical prediction. The resulting enhancement in the relative importance of the first terms can lead to anisotropic behavior $\delta S_{xx} \neq \delta S_{yy}$.
%\footnote{Note that $\delta \varepsilon_{xx} < \delta \varepsilon_{yy}$ since the former contains only the collisional contribution while the latter consists of both the band and the collisional contributions.}
This, however, needs to be verified experimentally
\footnote{Measurements of $S_{xx}$ in a ULSL have been reported in \cite{Taboryski95}, but to the knowledge of the present authors, no attempt to measure $S_{yy}$ has been made thus far.  We have also made measurements of the thermopower in ULSLs, and have found it difficult to obtain $S_{xx}$ and $S_{yy}$ correctly,  owing to the dominance of the $S_{xy}$ component and the tilting of the temperature gradient caused by the magnetic field \cite{Koike20TPCO,Endo19}}.
 
\section{Summary\label{SecSummary}}

To summarize, we have experimentally captured the CO in $R_{yx}$ of a ULSL, theoretically predicted some 30 years ago, by employing the measurement arrangement designed to efficiently pick out the extra component of $R_{yx}$ introduced by $V(x)$ and further by eliminating the parasitic component due to an unintentionally mixed $R_{xx}$ distinguishable as an even function of the magnetic field. The amplitude of the CO thus observed is found to be much smaller than the theoretical prediction. We have also deduced an asymptotic analytic expression for CO in the Hall conductivity $\delta \sigma_{yx}$ to facilitate the comparison between the theory and the experiment and to clarify the oscillation phase of $\delta \sigma_{yx}$. The smallness of $\delta \sigma_{yx}$ demonstrated in the present experiment suggests the possibility of considerable anisotropy in the Seebeck coefficient, contrary to the theoretical prediction. 

\begin{acknowledgments}
This work was supported by JSPS KAKENHI Grant Numbers JP20K03817 and JP19H00652.
\end{acknowledgments}

\appendix*

\section*{Appendix: Derivation of the Asymptotic Analytic Expressions\label{App}}
\setcounter{equation}{0}
In this Appendix, we describe the derivation of the asymptotic analytic expression $\sigma_{yx}^\mathrm{A}$ given by Eq.\ (\ref{sgmyxA}) from $\sigma_{yx}$ in Eq.\ (\ref{sgmyxnSum}).
Noting that $\lambda_N \alt 0.1$ for practical values of $V_0$, $a$, and $B$,  we obtain, up to $O(\lambda_N^2)$, 
\begin{widetext}
\begin{eqnarray}
\sigma_{yx} & \simeq & \displaystyle \frac{2e^2}{h}\ \sum_{N=0}^{\infty} \left(N+1\right) \int_{0}^{1}{d\xi\ \left\{f_{E_\mathrm{F}}\left(E_N\right)-f_{E_\mathrm{F}}\left(E_{N+1}\right)+\left[\frac{df_{E_\mathrm{F}}\left(E_N\right)}{dE}L_N\left(u\right)-\frac{df_{E_\mathrm{F}}\left(E_{N+1}\right)}{dE}L_{N+1}\left(u\right)\right]V_0e^{-\frac{u}{2}} \cos{\left(2\pi\xi\right)}\right\}} \nonumber \\
& & \displaystyle \times \left[1-2\lambda_N\cos{\left(2\pi\xi\right)}+3{\lambda_N}^2\cos^2{\left(2\pi\xi\right)}\right] \nonumber \\
& = & \displaystyle \sigma_{yx}^{(1)} + \sigma_{yx}^{(2)} \label{sgmyxnSumS}
\end{eqnarray}
with
\begin{equation}
\sigma_{yx}^{(1)} = \frac{2e^2}{h}\ \sum_{N=0}^{\infty} \left(N+1\right) \left[f_{E_\mathrm{F}}\left(E_N\right)-f_{E_\mathrm{F}}\left(E_{N+1}\right)\right]\left(1+\frac{3}{2}{\lambda_N}^2\right) \label{sgmyx1nSum}
\end{equation}
and
\begin{equation}
\sigma_{yx}^{(2)} = \frac{2e^2}{h}\ \sum_{N=0}^{\infty} \left(N+1\right)\left[-\frac{df_{E_\mathrm{F}}\left(E_N\right)}{dE}L_N\left(u\right)+\frac{df_{E_\mathrm{F}}\left(E_{N+1}\right)}{dE}L_{N+1}\left(u\right)\right]V_0e^{-\frac{u}{2}}\lambda_N, \label{sgmyx2nSum}
\end{equation}
where we performed the integration with respect to $\xi$. 
In the asymptotic limit of many filled Landau levels ($N \gg 1$), we can make the replacements, $e^{-u/2}L_N(u)$ $\rightarrow$ $(\pi^2 Nu)^{-1/4} \cos{(2\sqrt{Nu}-\pi/4)}$ and $e^{-u/2}L_{N}^{-1}(u)$ $\rightarrow$ $u(4\sqrt{\pi})^{-1}(Nu)^{-5/4}$$\left[4\sqrt{Nu}\cos{\left(2\sqrt{Nu}+\pi/4\right)-\sin{\left(2\sqrt{Nu}+\pi/4\right)}}\right]$, and take the continuum limit, $E_N \rightarrow E$, $\sum_{N}^\infty$ $\rightarrow$ $\int_{\hbar \omega_\mathrm{c}/2}^\infty dE/(\hbar\omega_\mathrm{c})$. The latter can be replaced by $\int_{-\infty}^\infty dE/(\hbar\omega_\mathrm{c})$ at low temperatures. We further make an approximation $f_{E_\mathrm{F}}\left(E_N\right)-f_{E_\mathrm{F}}\left(E_{N+1}\right) \simeq \hbar\omega_\mathrm{c} (-\partial f/\partial E)|_{E_\mathrm{F}-\frac{\hbar\omega_c}{2}}$. 
By performing the energy integral, we get
\begin{equation}
\sigma_{yx}^{(1)} \simeq \nu\frac{e^2}{h}\left[ 1+\frac{3}{2}{\lambda_\mathrm{c}}^2-A\left(\frac{T}{T_{a\mathrm{H}}}\right)\frac{3}{2}{\lambda_\mathrm{c}}^2\sin{\left(r_\mathrm{c}+\delta_\mathrm{F}\right)} \right] \label{sgmyx1EX}
\end{equation}
and $\sigma_{yx}^{(2)} = \sigma_{yx}^{(2\mathrm{a})} + \sigma_{yx}^{(2\mathrm{b})}$ with
\begin{equation}
\sigma_{yx}^{(2\mathrm{a})} \simeq \left(\nu+1\right)\frac{e^2}{h}\sqrt{\frac{2}{\pi}}\frac{V_0}{\hbar\omega_c}\frac{\lambda_\mathrm{c}}{\sqrt{r_\mathrm{c}^-}}\ \left[A\left(\frac{T}{T_{a\mathrm{H}}^+}\right)\cos{\left(\frac{r_\mathrm{c}^+ + r_\mathrm{c}^-}{2}+\frac{\delta_\mathrm{F}^+}{2}\right)}-A\left(\frac{T}{T_{a\mathrm{H}}^\delta}\right)\sin{\left(\frac{\delta_\mathrm{F}^+}{2}+\frac{r_\mathrm{c}^+-r_\mathrm{c}^-}{2}\right)}\right] \label{sgmyx2aEX}
\end{equation}
and
\begin{equation}
\sigma_{yx}^{(2\mathrm{b})} \simeq \left(\nu-1\right)\frac{e^2}{h}\sqrt{\frac{2}{\pi}}\frac{V_0}{\hbar\omega_c}\frac{\lambda_\mathrm{c}}{\sqrt{r_\mathrm{c}}}\left[A\left(\frac{T}{T_{a\mathrm{H}}^-}\right)\cos{\left(r_\mathrm{c}^-+\frac{\delta_\mathrm{F}^-}{2}\right)}-\frac{1}{\sqrt{1+{r_\mathrm{c}}^2}}\right], \label{sgmyx2bEX}
\end{equation}
\end{widetext}
where $\sigma_{yx}^{(2\mathrm{a})}$ ($\sigma_{yx}^{(2\mathrm{b})}$) derives from the first (second) term in Eq.\ (\ref{sgmyx2nSum}), $\lambda_\mathrm{c} \equiv 4\sqrt{2/\pi} [V_0/ (\hbar\omega_c)] u \sqrt{(1+r_\mathrm{c}^2)/r_\mathrm{c}^5}$, $T_{a\mathrm{H}} \equiv T_a (r_\mathrm{c}^2+1)/(r_\mathrm{c}^2-1)$, $r_\mathrm{c}^\pm \equiv r_\mathrm{c} \sqrt{1 \pm \nu^{-1}}$, $\delta_\mathrm{F}^\pm \equiv 2 \cot^{-1}{r_\mathrm{c}^\pm}$, $T_{a\mathrm{H}}^+ \equiv T_a \cdot 2 \sqrt{1-\nu^{-2}}({r_\mathrm{c}^+}^2+1)/[\sqrt{1+\nu^{-1}}({r_\mathrm{c}^+}^2+1)+\sqrt{1-\nu^{-1}}({r_\mathrm{c}^+}^2-1)]$, $T_{a\mathrm{H}}^- \equiv T_a \cdot \sqrt{1-\nu^{-1}} ({r_\mathrm{c}^-}^2+1)/{r_\mathrm{c}^-}^2$,  and $T_{a\mathrm{H}}^\delta \equiv T_a \cdot 2 \sqrt{1-\nu^{-2}}({r_\mathrm{c}^+}^2+1)/[\sqrt{1+\nu^{-1}}({r_\mathrm{c}^+}^2+1)-\sqrt{1-\nu^{-1}}({r_\mathrm{c}^+}^2-1)]$.

Noting that $\nu$, $r_\mathrm{c}$, $r_\mathrm{c}^\pm$ $\gg 1$ in the range of the magnetic field where CO is observed, we may neglect the difference between $T_a$ and $T_{a\mathrm{H}}$, $T_{a\mathrm{H}}^\pm$ to a good approximation. We can also make an approximation $\lambda_\mathrm{c} \simeq 4\sqrt{2/\pi} [V_0/ (\hbar\omega_c)] u r_\mathrm{c}^{-3/2}$ to attain the definition of $\lambda_\mathrm{c}$ presented in the main text. With these approximations, Eq.\ (\ref{sgmyx1EX}) becomes equivalent to Eq.\ (\ref{sgmyxA1}) in the main text. It can also readily be found that $A(T/T_{a\mathrm{H}}^\delta) \simeq 1$ at the cryogenic temperatures where CO is observed. After approximating $\lambda_\mathrm{c} / \sqrt{r_\mathrm{c}^-}$ in Eq.\ (\ref{sgmyx2aEX}) by $\lambda_\mathrm{c} / \sqrt{r_\mathrm{c}}$ for the sake of simplicity, we expand $(\nu+1)$ and $(\nu-1)$ in Eqs.\ (\ref{sgmyx2aEX}) and (\ref{sgmyx2bEX}), respectively. Then we collect the terms containing (not containing) the factor $\nu$, which yields Eq.\ (\ref{sgmyxA21}) [Eq.\ (\ref{sgmyxA22})] by further using the approximations $r_\mathrm{c}^\pm \simeq r_\mathrm{c} (1 \pm \nu^{-1}/2)$ and $\delta_\mathrm{F}^\pm \simeq \delta_\mathrm{F}\mp \nu^{-1} r_\mathrm{c}/(1+{r_\mathrm{c}}^2)$ within the $\cos$ and $\sin$ terms. The factor $\mathrm{sgn}(B)$ is incorporated in Eqs.\  (\ref{sgmyxA22bg}) and (\ref{dsgmyxA22}) to ensure the antisymmetry with respect to $B$. [See the caveat to Eq.\ (\ref{sgmyxnSum}) in the main text.] 

We note in passing that the expression $\sigma_{yx} = (2 e^2/h) (N+1) (1+ 3{\lambda_N}^2/2)$ presented just below Eq. (28) in \cite{Peeters92} \footnote{We have found that a factor 2 was missing in Ref.\ \cite{Peeters92}, which we resumed here.} corresponds to the low temperature limit of $\sigma_{yx}^{(1)}$ in the present study. As mentioned in the main text, $\sigma_{yx}^{(1)}$ only accounts for roughly half of the CO in $\sigma_{yx}$ [see also Fig.\ \ref{sgmyxcalc}(a)].

\bibliography{ourpps,thermo,lsls,twodeg,antidots}

\end{document}